\documentclass[aps,pra,10pt,a4paper,reprint,nofootinbib]{revtex4-2}
\usepackage[utf8]{inputenc}
\usepackage[T1]{fontenc}
\usepackage{amsfonts}
\usepackage{amssymb}
\usepackage{amsmath}
\usepackage{amsthm}
\usepackage{graphics}
\usepackage{graphicx}
\usepackage{braket}
\usepackage[colorlinks=true,linkcolor=blue,
urlcolor=blue,citecolor=red]{hyperref}

\newtheorem{definition}{Definition}

\begin{document}
\title{Separability and entanglement in superpositions of quantum states}

\author{Saronath Halder}
\affiliation{Harish-Chandra Research Institute, A CI of Homi Bhabha National Institute, Chhatnag Road, Jhunsi, Allahabad 211 019, India}

\author{Ujjwal Sen}
\affiliation{Harish-Chandra Research Institute, A CI of Homi Bhabha National Institute, Chhatnag Road, Jhunsi, Allahabad 211 019, India}

\begin{abstract}
It is known that probabilistically mixing an arbitrary pair of pure quantum states, one of which is entangled and the other product, in any bipartite quantum system, one always obtains an entangled state, provided the entangled state of the pair appears with a nonzero probability. On the other hand, if we consider any superposition of the same pair, with a nonzero amplitude for the entangled state of the pair, the output state may not always be entangled. Motivated by this fact, in this work, we study the superpositions of a pure entangled state and a pure product state, when the amplitudes corresponding to the states appearing in any superposition are nonzero. We show, in particular, that all such superpositions produce only entangled states if the initial entangled state has Schmidt rank three or higher. Again, superposing a pure entangled state and a product state cannot lead to product states only, in any bipartite quantum system. These lead us to define conditional and unconditional inseparabilities of superpositions. These concepts in turn are useful in quantum communication protocols. We find that conditional inseparability of superpositions help in identifying strategies for conclusive local discrimination of shared quantum ensembles. We also find that the unconditional variety leads to systematic methods for spotting ensembles exhibiting the phenomenon of more nonlocality with less entanglement and two-element ensembles of conclusively and locally  indistinguishable shared quantum states.
\end{abstract}
\maketitle

\section{Introduction}\label{sec1}
Quantum entanglement \cite{Horodecki09-1, Guhne09, Das17} is regarded as an important resource because it finds applications in several information processing protocols, like quantum teleportation \cite{Bennett93, Bouwmeester97}, quantum dense coding \cite{Bennett92, Mattle96, Bruss04, Prabhu13, Prabhu13-1, Das15}, and quantum key distribution \cite{Bennett14,Ekert91}. Therefore, it is important to understand the origin and characteristics of such a resource. Entanglement in quantum states of shared systems is a result of the superposition principle of quantum physics. Arbitrary superpositions, however, do not lead to  entangled states of the corresponding shared physical system. For example, if we consider superpositions of a pure entangled state and a product state, then we may not always get entanglement as output. In this context, we mention that if we mix any pure entangled state with an arbitrary product state with nonzero probabilities then we always get an entangled state \cite{Horodecki03-1}. Motivated by this fact, in this work we consider the superpositions of a pure entangled state and a product state, when the amplitudes (i.e., superposition coefficients) corresponding to the (normalized) states appearing in any superposition are nonzero. 

The question of entanglement of superposed quantum states has already been posed in the literature. The subsequent studies have, as far as we know, been always quantitative, providing important bounds on the amount of entanglement generated in different types of superpositions,  considering different entanglement measures \cite{Linden06, Gour07, Yu07, Ou07, Niset07, Cavalcanti07, Song07, Gour08, Osterloh08, Akhtarshenas11, Parashar11, Ma14}. We, however, wish to study the problem qualitatively - answering only whether a given superposition of quantum states of a shared system is entangled or not. In particular, we ask when the superpositions of a given pair of a pure entangled state and a product state can produce entangled states, provided that the coefficients corresponding to the states appearing in any superposition are nonzero. A quantitative study of course provides the qualitative answer, but we find that in several cases where the quantitative solution is missing, one can still extract  qualitative results. This is one of the reasons why we focus on a qualitative study in this work. Furthermore, we provide several applications of such a study.

In discussions about robustness of entanglement, one often considers the mixtures of an entangled state with a separable state, i.e., convex combination of the states, where the latter is deemed as ``noise'' \cite{Lewenstein98, Vidal99-1, Du00, Ishizaka00, Verstraete01, Simon02, Harrow03, Steiner03, Horodecki03-1, Brandao05, Bandyopadhyay05, Cavalcanti06, Bandyopadhyay08, Guhne08, Chaves10, Benatti12, Halder19-1}. By considering such a mixture, one examines how robust the given entangled state is, against the chosen noise. In other words, we ask how much mixing of the noise does the entangled state tolerate, so that the newly produced state remains entangled. Sometimes, it is possible to find a separable state for a given entangled state, such that any mixture of the states  produce entangled states, provided that the initial entangled state appears with nonzero probability in any of the mixtures. In such a scenario,  we can say that the entangled state is {\it unconditionally robust} in the direction of that separable state \cite{Ishizaka00, Verstraete01, Horodecki03-1, Bandyopadhyay08, Halder19-1}. As mentioned earlier, for an arbitrary pair of a pure entangled state and a product state, any mixture of the states with nonzero probabilities produces entangled states only \cite{Horodecki03-1}. Therefore, all pure entangled states are unconditionally robust in the direction of all product states. Following this notion of unconditional robustness, we ask whether it is possible to find product states, such that any superposition of a product state with a given pure entangled state is always entangled when the coefficients corresponding to the states appearing in any superposition are nonzero. If it is possible to produce such an instance, then we refer to the phenomenon as {\it unconditional inseparability of superpositions}.

In general, we find that for all pairs of a pure entangled state and a product state, arbitrary superpositions lead to entangled states only, when the initial entangled state has Schmidt rank three or above and the coefficient corresponding to it is nonzero in any superposition. Clearly, for bipartite quantum systems, none of whose constituents is a qubit, all pure entangled states are ``unconditionally superposition robust'' in the direction of all product states, when the initial entangled state has Schmidt rank three or above. This, therefore, provides a parallel scenario in the context of superpositions, with respect to the result stated before from Ref.~\cite{Horodecki03-1} in the complementary context of mixtures. The opposite problem is to uncover instances where superpositions produce product states. We find that in arbitrary bipartite quantum systems, a pure entangled state and a product state cannot lead, via superpositions, to product states only. We refer to this phenomenon as {\it conditional inseparability of superpositions}. These results about superposing a pure entangled state and a product state in arbitrary bipartite quantum systems are given in Sec.~\ref{sec2}. Thereafter, we provide some detailed discussions related to the two-qubit system in Sec.~\ref{sec2a}. 

In Sec.~\ref{sec3}, we discuss about the applications of the above findings. For example, we show that the phenomenon of {\it unconditional inseparability of superpositions} finds application in the context of ``nonlocality''  associated with the problem of state discrimination  under local quantum operations and classical communication (LOCC) \cite{Bennett99-1, Walgate00, Virmani01, Ghosh01, Groisman01, Terhal01, Eggeling02, Walgate02, Ghosh02, Horodecki03, Chen03, Chen03-1, Badziag03, Horodecki04, Ghosh04, Rinaldis04, Fan04, Ghosh05, Nathanson05, Watrous05, Niset06, Hayashi06, Sen(De)06, Horodecki07, Duan07, Feng09, Matthews09, Bandyopadhyay10-1, Bandyopadhyay11, Yu12, Bandyopadhyay12, Yang13, Childs13, Zhang14, Zhang15, Xu16, Croke17, Lami18, Halder18, Halder19, Halder20-1, Lami21}. We find that the phenomenon of {\it conditional inseparability of superpositions} can also have important applications in the context of local quantum state discrimination problems. We then identify a class of unextendible entangled bases, which we refer to as ``\(r\)-UEBs'', and prove that they can contain no state which is conclusively locally identifiable with nonzero probability.

Some discussions on the results, including a comparison with the quantitative results already known in the literature, are presented in Sec.~\ref{sec4}. 

\section{Superposing an entangled state and a product state}\label{sec2}
We provide two results in this section for the general case, that is for pure states of arbitrary bipartite dimensions. Both concern the entanglement or its absence in a superposition of a pure entangled state and a product state. We will use the concept of the \emph{Schmidt rank} of bipartite pure quantum states, which is defined as the number of nonzero (Schmidt) coefficients in a Schmidt decomposition of the (bipartite pure) state \cite{Horodecki09-1, Guhne09, Das17}. However, before we provide the theorems, we formally define the concepts of conditional and unconditional inseparabilities of superposition.

\begin{definition}{[Conditional and unconditional inseparabilities of superposition]} If it is possible to find a product state, such that any superposition of the product state with a given pure entangled state is always entangled, when the coefficients corresponding to the states appearing in any superposition are nonzero, then we say that the initial entangled state is unconditionally robust in the direction of the product state and we refer to the phenomenon as an instance of unconditional inseparability of superpositions. Otherwise, it is an instance of conditional inseparability of superpositions. \end{definition}

\noindent\textbf{Theorem 1.}
\emph{Any nontrivial superposition of an arbitrary entangled pure state  and an arbitrary  product state of a bipartite quantum system is entangled, provided the initial entangled state has Schmidt rank three or higher.}\\

\noindent \textbf{Remark 1.} Among bipartite quantum systems, barring those for which a local dimension is two (i.e., barring \(\mathbb{C}^2 \otimes \mathbb{C}^d\) systems), a pure state can have Schmidt rank three or higher. 

\noindent \textbf{Remark 2.} By a ``nontrivial'' superposition of two pure quantum states, we imply that the initial states are not included in the discussion. I.e., if we consider any superposition of two states $\ket{e}$ and $\ket{p}$ as $a_1\ket{e}+a_2\ket{p}$, then the coefficients $a_1$ and $a_2$ corresponding to the states $\ket{e}$ and $\ket{p}$ must be nonzero. If one of them is zero then after superposition, we get back one of the initial states, i.e., $\ket{e}$ or $\ket{p}$. We do not include these ``trivial'' cases here.

\noindent \textbf{Remark 3.} We
therefore find that for two qutrits and higher local dimensions, all pure entangled states of Schmidt rank three or higher are ``unconditionally superposition robust'' in the direction of all product states. The complementary situation, where a pure entangled state is mixed with a product state, always leads to an entangled state, is known from Ref.~\cite{Horodecki03-1}.

\begin{proof}
Let $\ket{e}$ be an entangled state with Schmidt rank $\geq3$ and let $\ket{p}$ be a product state. If possible, let $a_1\ket{e} + a_2\ket{p} = \ket{p^\prime}$ be a product state, where \(a_1\), \(a_2\) are nonzero complex numbers. So, $a_1\ket{e} = \ket{p^\prime} - a_2\ket{p}$. The vector  $a_1\ket{e}$ is not normalized, but it is possible to define the Schmidt rank of this element, which is exactly equal to that of $\ket{e}$, i.e., $\geq3$. Similarly, if we define the Schmidt rank for the vector $\ket{p^\prime} - a_2\ket{p}$, then it can be shown that it is not greater than two, giving us a contradiction, proving that the initial assumption of \(|p^\prime\rangle\) being a product state is not true. We are therefore left with proving that $\ket{p^\prime} - a_2\ket{p}$ has Schmidt rank \(\leq 2\) (where \(|p\rangle\) and \(|p^\prime\rangle\) are two product states and \(a_2\) is a nonzero complex number), which we do now. This can be understood in the following way. We assume that  $\ket{p} = \ket{\alpha}\ket{\beta}$ and $\ket{p^\prime} = \ket{\alpha^\prime}\ket{\beta^\prime}$. If the states on any of the sides of the bipartite system are linearly dependent, i.e., if either \(|\alpha\rangle\) and  \(|\alpha^\prime\rangle\) are linearly dependent or \(|\beta\rangle\) and  \(|\beta^\prime\rangle\) are so (or both), then $\ket{p^\prime}- a_2\ket{p}$ can be written in tensor product form, and so will have unit Schmidt rank. Let therefore $\ket{\alpha}$ and $\ket{\alpha^\prime}$ be linearly independent, and  similarly, let $\ket{\beta}$ and $\ket{\beta^\prime}$ be also so. Then, $\ket{\alpha^\prime}$ can be written as $b_0\ket{\alpha}+b_1\ket{\alpha^\perp}$, where \(\ket{\alpha}\) and \(\ket{\alpha^\perp}\) are orthogonal to each other, and where \(b_0\) and \(b_1\) are complex numbers. Then we can rewrite  $\ket{p^\prime}- a_2\ket{p}$ as $\ket{\alpha}(b_0\ket{\beta^\prime}-a_2\ket{\beta})+b_1\ket{\alpha^\perp}\ket{\beta^\prime}$. Tracing out the first party, the reduced density matrix of the second is a convex combination (probabilistic mixture, but possibly not normalized to unit probability) of the vectors \(b_0\ket{\beta^\prime}-a_2\ket{\beta}\) and \(b_1\ket{\beta^\prime}\). This matrix cannot have more than two nonzero eigenvalues, which implies that the Schmidt rank of \(\ket{p^\prime}- a_2\ket{p}\) is \(\leq 2\). This completes the proof.
\end{proof}

\noindent \textbf{Generalization of Theorem 1.} \emph{Let us consider an entangled state $\ket{e}$ of Schmidt rank, $r\geq3$, and also $r-2$ product states, \(\ket{p_1}\), \(\ket{p_2}\), \(\ldots\), \(\ket{p_{r-2}}\). Then the state $\ket{\psi} = a_0\ket{e}+a_1\ket{p_1}+a_2\ket{p_2}+\ldots+a_{r-2}\ket{p_{r-2}}$ is always entangled, where $a_i$ are complex numbers such that $\langle\psi|\psi\rangle=1$ and \(a_0 \ne 0\).}

\begin{proof}Consider the element $\ket{e_1^\prime} := a_0\ket{e}+a_1\ket{p_1} $. We claim that $\ket{e_1^\prime}$ cannot have Schmidt rank less than $r-1$. This follows from the following  contradiction. Suppose, $\ket{e_1^\prime}$ has Schmidt rank $\leq r-2$. So, $\ket{e_1^\prime}$ = $\ket{0}\ket{0^\prime}+\ket{1}\ket{1^\prime}+\cdots+\ket{(l-1)}\ket{(l-1)^\prime}$, where $l\leq r-2$, and where we have ignored the normalization of the constituent Schmidt kets and the overall superposition. Now, $a_0\ket{e} = \ket{e_1^\prime}-a_1\ket{p_1}$. We can take $\ket{p_1} = \ket{\alpha}\ket{\beta}$, where \(\ket{\alpha} = c_0\ket{0}+c_1\ket{1}+c_2\ket{2}+\cdots+c_{l-1}\ket{l-1}+c_{l}\ket{l^\perp}\) with $\ket{l^\perp}$ being orthogonal to the mutually orthogonal kets, $\ket{0}, \ket{1}, \ket{2}, \ldots, \ket{l-1}$. So, $\ket{e_1^\prime}-a_1\ket{p_1}$ can be written as $\ket{0}(\ket{0^\prime}-a_1c_0\ket{\beta})+\ket{1}(\ket{1^\prime}-a_1c_1\ket{\beta})+\cdots+\ket{(l-1)}(\ket{(l-1)^\prime}-a_1c_{l-1}\ket{\beta})-a_1c_l\ket{l^\perp}\ket{\beta}$. This cannot have Schmidt rank $>r-1$ because tracing out the first party lands us in a state that has support on a space spanned by $\leq r-1$ kets. But $a_0\ket{e}$ has Schmidt rank $r$. Clearly, the Schmidt rank of $a_0\ket{e}$ cannot be equal to that of $\ket{e_1^\prime}-a_1\ket{p_1}$. Thus, $\ket{e_1^\prime}$ cannot have Schmidt rank less than $r-1$. In a similar fashion, we can prove that the element $a_0\ket{e}+a_1\ket{p_1}+a_2\ket{p_2}$ cannot have Schmidt rank less than $r-2$, and finally, $\ket{\psi}$ cannot have Schmidt rank less than \(r-r+2 = 2\). Note that in the above argument, if we take $c_l=0$ then anyway, the contradiction will occur. This completes the proof of the generalization of Theorem 1. 
\end{proof}
The next result looks at a scenario that is  complementary to the one answered in the foregoing theorem.\\

\noindent 
\textbf{Theorem 2.} \emph{There does not exist a pair of a pure entangled state and a product state such that any superposition of the states  produce a product state, when the coefficients corresponding to the initial states in the superpositions are nonzero.}

\begin{proof}
Theorem 1 implies that the current proof needs to be done only for entangled states of Schmidt rank two. However, we provide a general proof here. Let \(|e\rangle\) be an entangled state and \(|p\rangle\) a product of an arbitrary bipartite quantum system. Consider now the superposition \(|\psi\rangle = \epsilon |e\rangle + \sqrt{1 - |\epsilon|^2} |p\rangle\), for arbitrary nonzero and non-unit complex number \(\epsilon\) such that \(\langle \psi | \psi \rangle = 1\). Let us consider the metric \(d(|\phi_1\rangle, |\phi_2\rangle) = \sqrt{1- |\langle \phi_1 | \phi_2 \rangle|^2}\) on the tensor-product Hilbert space corresponding to the bipartite quantum system under consideration. The set of states \(|\psi\rangle\) generated by varying \(\epsilon\), now considered to be real and \(\in (0,1)\), can be understood as a continuous ``line segment'' connecting the  ``points'' \(|e\rangle\) and \(|p\rangle\) on the joint Hilbert space. Now since product states form a closed set in this Hilbert space with respect to the metric \(d\), there is always an \(\epsilon_1 >0\) such that \(|\psi\rangle\) is entangled for all \(\epsilon \in [\epsilon_1,1]\). 
\end{proof}

The above results, and especially the remark after Theorem 1, clearly underline the importance of considering bipartite systems of low dimensions for analyzing entanglement in superpositions. And we consider the two-qubit case in detail in the following section.
 
\section{Two-qubit systems: Unconditional inseparability of superpositions}\label{sec2a}
Theorem 1 is void for two-qubit systems (actually for all \(\mathbb{C}^2 \otimes \mathbb{C}^d\) systems). There are no pure states of two qubits that are of Schmidt rank three or higher. The landscape is richer here than in the higher dimensions considered in Theorem 1, and a pure entangled  state, when superposed with a product state, can lead to entangled as well as product states, in the case of two-qubit systems. 

Let us first discuss the cases when the output (i.e., the superposed state) is a product state. An example is obtained by superposing \((|00\rangle + |11\rangle)/\sqrt{2}\) and \(|11\rangle\) with suitable coefficients, so that the output is \(|00\rangle\). If we consider $a_1(|00\rangle + |11\rangle)/\sqrt{2}+a_2|11\rangle$, then we can take $a_1=\sqrt{2}$ and $a_2=-1$ to get $|00\rangle$. In this example, the input entangled and product states are nonorthogonal. This however is not a necessity, and we now give an example of a superposition of a pure entangled state and an orthogonal product state, such that the output is a product. This can be performed systematically by the usual method of solving the eigenvalue equation of a local density matrix of the two-qubit pure state. As we will see in the succeeding section, this exercise can be of crucial importance in identifying ensembles of locally indistinguishable shared states.\\

\noindent 
\emph{Example 1.} Consider the two-qubit entangled state, $|e_1\rangle = 2/\sqrt{5}\ket{00}+1/\sqrt{5}\ket{11}$, written in Schmidt decomposition. We ask the following question: Is there a  product state that is orthogonal to this entangled state, and for which a superposition of the product state  with the given entangled state can produce a product state? To answer this question, we write down an arbitrary superposition of the two states, apply the condition of orthogonality, and solve the eigenvalue equation of a local density matrix of the superposed state. Some algebra leads us to the product state \(|p_1\rangle = (2/\sqrt{5}\ket{0}-1/\sqrt{5}\ket{1})(1/\sqrt{17}\ket{0}+4/\sqrt{17}\ket{1})\), for which the superposition \(3/\sqrt{26}|e_1\rangle + \sqrt{17/26}|p_1\rangle\) is a product state. 

We therefore see that a two-qubit pure entangled state can superpose with a two-qubit product state to create a product state. The same pairs will always lead to at least some entangled states when other superposition coefficients are considered, as guaranteed by Theorem 2. Such pairs are what we can refer to as leading to {\it conditional inseparability of superpositions}. However, we are now interested to discuss about  {\it unconditional inseparability of superpositions}, which refers to a set of pure quantum states, any superposition of which can only produce an entangled state. We provide now two examples of such unconditional inseparability of superpositions, where we focus only on sets which consist of a two-qubit pure entangled state and a two-qubit product state. \\

\noindent 
{\it Example 2.} An arbitrary two-qubit pure entangled state can be written, in Schmidt form, as $a_1\ket{00}+a_2\ket{11}$, where $a_1, a_2$ are (nonzero) positive real numbers such that $a_1^2+a_2^2=1$. For varying \(a_1\) and \(a_2\), a two-dimensional subspace is spanned, whose orthogonal complement contains the product states \(|01\rangle\) and \(|10\rangle\). One can check that any superposition of \(a_1\ket{00}+a_2\ket{11}\) with \(|01\rangle\) or \(|10\rangle\) is always entangled - unconditional inseparability of superpositions. This fact can be generalized to higher dimensions, but Theorem 1 reports an even better generalization there. \\

\noindent 
{\it Example 3.} Consider now the maximally entangled state, $(\ket{00}-\ket{11})/\sqrt{2}$, and the (orthogonal) product state, \(|++\rangle\), where \(|+\rangle = (|0\rangle + |1\rangle)/\sqrt{2}\). Superpositions of these two states produce entangled states only. Therefore, this pair provides  another example of unconditional inseparability of superpositions.\\

\noindent \textbf{Unconditional inseparability and nonorthogonality.} We will see in the following section that superpositions of a pure entangled  state and an \emph{orthogonal} product state is of importance in local state discrimination tasks. However, it is possible to find examples of unconditional inseparability of superpositions even in a pair consisting of an entangled and a product state which are nonorthogonal. Consider the entangled state, $\ket{e} = a_1\ket{00}+a_2\ket{11}$, where $a_1,a_2$ are nonzero Schmidt coefficients and $a_1^2+a_2^2=1$. We try to figure out a possible structure of a product state $\ket{p}$, not necessarily orthogonal to $\ket{e}$, such that $\ket{\psi}=e\ket{e}+p\ket{p}$ is entangled  for all nonzero values of $e$ and $p$, with $\langle\psi|\psi\rangle=1$. Obviously, $\ket{p}$ has the structure, $\alpha\ket{e}+\beta\ket{e^\perp}$, with $|\alpha|^2+|\beta|^2=1$, and where \(\ket{e^\perp}\) is orthogonal to \(|e\rangle\). We  assume that $\ket{e^\perp}$ is an entangled state which has the structure, $\ket{e^\perp} = a_3\ket{01}+a_4\ket{10}$, $a_3,a_4$ are nonzero Schmidt coefficients and $a_3^2+a_4^2=1$. Based on the values of $a_1, a_2, a_3, a_4$, the values of $\alpha,\beta$ will be fixed for which $\ket{p}$ will be a product state. We now prove that $\alpha, \beta$ must be real and they form a  unique pair, for \(\alpha\ket{e}+\beta\ket{e^\perp}\) to be a product state. To prove this, we consider the superposition $\cos{\frac{\theta}{2}}(a_1\ket{00}+a_2\ket{11})+e^{i\phi}\sin{\frac{\theta}{2}}(a_3\ket{01}+a_4\ket{10})$. This state is a product state iff $|\cos^2{\frac{\theta}{2}}a_1a_2-e^{i2\phi}\sin^2{\frac{\theta}{2}}a_3a_4|=0$. This implies that $\cos^2{\frac{\theta}{2}}a_1a_2-\cos{2\phi}\sin^2{\frac{\theta}{2}}a_3a_4=0$ and $\sin{2\phi}\sin^2{\frac{\theta}{2}}a_3a_4 = 0$. The second equation implies that $\sin{2\phi}=0$, i.e., $\phi=\frac{n\pi}{2}$, with $n$ being an integer. Putting this value of $\phi$ in the equation $\cos^2{\frac{\theta}{2}}a_1a_2-\cos{2\phi}\sin^2{\frac{\theta}{2}}a_3a_4=0$, we get $\tan^2{\frac{\theta}{2}} = \frac{(-1)^n a_3a_4}{a_1a_2}$. But $\tan^2{\frac{\theta}{2}}$ cannot be negative, and thus we have to take $(-1)^n = 1$, i.e., $n$ must be an even number. We now set $\frac{a_3a_4}{a_1a_2}=k$, a constant. Thus, we get $\alpha=\cos{\frac{\theta}{2}}=\sqrt{\frac{k}{k+1}}$ and $\beta=\sin{\frac{\theta}{2}}=\sqrt{\frac{1}{k+1}}$. Clearly therefore,  $\alpha, \beta$ are real, and they form a unique pair. Next, we substitute  $\ket{p}$ = $\alpha\ket{e}+\beta\ket{e^\perp}$ in the expression of $\ket{\psi} = e\ket{e}+p\ket{p}$, to get $\ket{\psi}$ = $(e+p\alpha)\ket{e}+p\beta\ket{e^\perp}$. We now compare this expression with the expression of $\ket{p}$ = $\alpha\ket{e}+\beta\ket{e^\perp}$. We have just seen that there is a unique combination of coefficients for a linear combination of \(|e\rangle\) and \(|e^\perp\rangle\) to be a product state, and that is the state \(|p\rangle\). So, all other linear combinations in \(\ket{\psi} = (e+p\alpha)\ket{e}+p\beta\ket{e^\perp} = e\ket{e}+p\ket{p}\) form entangled states. Moreover, \(\langle e|p\rangle = \alpha = \sqrt{\frac{k}{k+1}} \ne 0\), so that \(|e\rangle\) and \(|p\rangle\) are not orthogonal. We therefore have identified a class of pairs, each consisting of an entangled state and a nonorthogonal product state that provide  unconditional inseparability of superpositions.

\section{Applications}\label{sec3}
In this section, we demonstrate that the concepts and corresponding examples that we discussed in the preceding sections can be useful in certain quantum communication tasks. 

\subsection{Unextendible entangled basis and conclusive local discrimination of quantum states}\label{subsec3}
We begin with  the definition of an unextendible entangled basis~(UEB)~\cite{Bravyi11, Chen13, Li14, Wang14, Nan15, Nizamidin15, Guo16, Zhang16-3, Wang17-1, Zhang18-1, Zhang18-2, Liu18, Song18, Zhao20, Guo14, Han18, Shi19, Yong19, Wang19, Chakrabarty12,Chen13-1, Guo15-1, Zhang17-4, Halder20}.

\begin{definition}{[Unextendible entangled basis]} An unextendible entangled basis is a set of mutually orthonormal entangled states of a composite Hilbert space such that there are no entangled states in the orthogonal complement of their span. \end{definition}

We note here that the span of the entangled states within a UEB must not be equivalent to the whole given Hilbert space, for the concept of UEB to be nontrivial. While the definition has been generalized to the multiparty case, we focus on the bipartite case only. We consider a special type of unextendible entangled basis in $\mathbb{C}^{d_1}\otimes\mathbb{C}^{d_2}$, $d_1, d_2>2$, as given by the succeeding definition. 

\begin{definition}{[$r$-UEB]} Let \(r\) be a positive integer and $r\geq3$. We call an unextendible entangled basis as an \(r\)-UEB if all elements of that UEB has a Schmidt rank \(r\) or higher, and there is at least one element of Schmidt rank \(r\).\end{definition}

For $r$-UEBs, we can present Theorem 3, given below. We note here that ``conclusive identification'' of each state, in a given set of states,  with some nonzero probability is required for ``conclusive distinguishability'' of a given set. And to identify a quantum state, drawn from the given set of states, conclusively under LOCC, it is necessary and also sufficient to find a product state which has nonzero overlap with the considered state but the product state must have zero overlap with the other states of the set \cite{Chefles04}.\\

\noindent \textbf{Theorem 3.} \emph{An \(r\)-UEB contains no state which is conclusively locally identifiable.}

\begin{proof} 
If we draw any state from the given UEB, it is not possible to produce a product state by taking any superposition of the drawn state and the product states from the complementary subspace. This follows from Theorem 1.  Suppose that the drawn state is $\ket{e}$, and let $\{\ket{p_1}, \ket{p_2}, \ldots, \ket{p_n}\}$ be a (product) basis for the complementary subspace. Let $\ket{\phi} = a_0\ket{e}+a_1\ket{p_1}+\ldots+a_n\ket{p_n}$, where $a_i$ are nonzero complex numbers such that $\langle\phi|\phi\rangle=1$. We can rewrite $\ket{\phi}$ as $a_0\ket{e}+a^\prime\ket{p^\prime}$, where $a^\prime = \sqrt{|a_1|^2+|a_2|^2+\cdots+|a_n|^2}$ and where $\ket{p^\prime}=(1/a^\prime)(a_1\ket{p_1}+\cdots+a_n\ket{p_n})$ is another product state (by the definition of UEB). Therefore, $\ket{\phi}$ is an entangled state (by Theorem 1). So, it is not possible to find any product state which is nonorthogonal to the drawn state but orthogonal to the rest of the states of the UEB. 

It is also important to mention the following: The state $|\phi\rangle = a_0^\prime|e\rangle + a_1^\prime|e^\prime\rangle$, $|a_0^\prime|^2+|a_1^\prime|^2=1$, can be a product state but it will be nonorthogonal to $|e^\prime\rangle$ also, where $|e\rangle,~|e^\prime\rangle$ are different states drawn from the UEB. This is not desired because to identify the state $|e\rangle$ unambiguously with some nonzero probability, we need a $|\phi\rangle$ which must be nonorthogonal to $|e\rangle$ but orthogonal to $|e^\prime\rangle$. This is the criterion we get from \cite{Chefles04}.

These complete the proof. 
\end{proof}

The proof of Theorem 3 also implies that given a complete orthonormal basis, whose states are all entangled,  no state of that basis can be conclusively locally identified. This is because there is no room to find a product state which is nonorthogonal to the drawn state from the given basis. See Ref.~\cite{Horodecki03} in this regard. Note also that when an incomplete entangled basis is given, the conclusive local identifiability is not that obvious.

To provide an example of an \(r\)-UEB, considered in Theorem 3,  we construct a \(3\)-UEB in $\mathbb{C}^4\otimes\mathbb{C}^4$. We first identify nine entangled states of Schmidt rank 3, viz., the states of the set, \{$\ket{\psi_1}$, $\ket{\psi_2}$, $\ket{\psi_3}$\}, belonging to the subspace spanned by the states in \{$\ket{00}$, $\ket{11}$, $\ket{22}$\}, the states in \{$\ket{\psi_4}$, $\ket{\psi_5}$, $\ket{\psi_6}$\}, belonging to the subspace spanned by \{$\ket{01}$, $\ket{12}$, $\ket{20}$\}, and the states in \{$\ket{\psi_7}$, $\ket{\psi_8}$, $\ket{\psi_9}$\}, belonging to the subspace spanned by \{$\ket{02}$, $\ket{10}$, $\ket{21}$\}. In fact, these nine states can be orthogonal to each other. Next, we consider the product states \(\ket{03}\), \(\ket{13}\), \(\ket{23}\), \(\ket{30}\), \(\ket{31}\), \(\ket{32}\), \(\ket{33}\). These product states are also orthogonal to the previously mentioned nine entangled states. We now present the final basis in $\mathbb{C}^4\otimes\mathbb{C}^4$. It consists of the states, $\frac{1}{\sqrt{2}}(\ket{\psi_1}\pm\ket{03})$, $\frac{1}{\sqrt{2}}(\ket{\psi_2}\pm\ket{13})$, $\frac{1}{\sqrt{2}}(\ket{\psi_3}\pm\ket{23})$, $\frac{1}{\sqrt{2}}(\ket{\psi_4}\pm\ket{30})$, $\ket{\psi_5}$, $\ket{\psi_6}$, $\ket{\psi_7}$, $\ket{\psi_8}$, $\ket{\psi_9}$, $\ket{31}$, $\ket{32}$, $\ket{33}$. Clearly, the first thirteen states form a UEB in $\mathbb{C}^4\otimes\mathbb{C}^4$ with the property that the entangled states have Schmidt rank three. This UEB constitutes an example of an \(r\)-UEB (for \(r=3\)) as used in  Theorem 3.

The UEBs of Theorem 3, exhibit a stronger form of nonlocality, compared to those sets of states which are not perfectly locally distinguishable but conclusively locally distinguishable. These UEBs are also more nonlocal compared to those sets of states, a few states of which are conclusively locally identifiable but all of them are not.

We should note here that the above type of basis is not possible in $\mathbb{C}^2\otimes\mathbb{C}^d$. In that case, Schmidt rank $\geq3$ is not possible for any pure state. Particularly, for two qubits, it is possible to construct UEBs of cardinality three only \cite{Halder20}, at least one state of which is conclusively locally identifiable \cite{Bandyopadhyay09}.\\

\noindent \textbf{Partially entangled subspaces.} A ``partially entangled subspace'' contains both entangled and product states but it is deficient in product states, so that  it is not possible to find a product basis for the subspace. Such a partially entangled subspace can be obtained by using an \(r\)-UEB. Consider any (entangled) state drawn from an \(r\)-UEB. The drawn state along with the product basis for the complementary subspace of the \(r\)-UEB, produce a partially entangled subspace. We note that the product state deficit is due to the fact that the drawn state will superpose with all other states to form only entangled states (by Theorem 1), thus always blocking at least one dimension in which there is only an entangled state, so that no basis for the subspace can be formed using only product states. An interesting feature of a partially entangled subspace is that all full-rank states associated with the subspace is entangled, as they violate the range criterion. This is directly due to the product state deficit in the subspace.\\

\noindent\textbf{Generalization of Theorem 3.} Following the generalization of Theorem 1 along with Theorem 3, we can consider another type of UEB in a bipartite system \(\mathbb{C}^{d_1} \otimes \mathbb{C}^{d_2}\), where the entangled states have Schmidt rank at least $r$ and the complementary subspace contains states of Schmidt rank $\leq (r-2)$. Then, no state of the UEB can be conclusively identified by LOCC.

\subsection{Strategies for conclusive local discrimination}
Suppose that we are given a set of three two-qubit pure mutually-orthogonal entangled states $\ket{\Psi_1}$, $\ket{\Psi_2}$, and $\ket{\Psi_3}$, such that the (unique) state which is orthogonal to these states, is a product state. Such a set cannot be perfectly distinguished by separable measurements \cite{Duan09, Halder20}, which is a strict superset of the set of LOCC-based measurements \cite{Bennett99-1}. Nevertheless, it is possible to identify, by LOCC, at least one state of the set conclusively with some nonzero probability \cite{Bandyopadhyay09}. We remember here the definitions of conclusive distinguishability and identification, given in Sec.~\ref{subsec3}. And we reiterate that  to identify a quantum state, drawn from a given set of states, conclusively under LOCC-based measurement strategies, it is necessary and sufficient to find a product state which has nonzero overlap with the considered state but the product state must have zero overlap with the other states of the set \cite{Chefles04}.  In a practical scenario, it will be important to know the form of such product states to prepare a measurement setup for the conclusive identification. Now, we go back to the given set of three states. Suppose, the state $\ket{\Psi_1} = 2/\sqrt{5}\ket{00}+1/\sqrt{5}\ket{11}$, and we want to identify this state conclusively, by LOCC, with some nonzero probability. We also assume that the product state, which is orthogonal to the states $\ket{\Psi_i}$, $\forall i =1,2,3$, is given by \(|\Phi\rangle = (2/\sqrt{5}\ket{0} - 1/\sqrt{5}\ket{1})  (1/\sqrt{17}\ket{0}+4/\sqrt{17}\ket{1})\). Clearly, to identify the state $\ket{\Psi_1}$ conclusively under LOCC with some nonzero probability, we have to find out a product state by taking a superposition of $\ket{\Psi_1}$ and $\ket{\Phi}$. This can be found from {\it Example 1} in the preceding section.  

\subsection{More nonlocality with less entanglement}
Consider a pair consisting of a pure entangled normalized state $\ket{\widetilde{\Psi}_1}$ and an orthogonal product normalized state $\ket{\widetilde{\Phi}}$ in a two-qubit system, such that a nontrivial superposition of the pair is always entangled. So for arbitrary complex numbers, \(a_1\) and \(a_2\), with \(a_1,a_2 \ne 0\) and \(|a_1|^2+|a_2|^2=1\), $a_1\ket{\widetilde{\Psi}_1}+a_2\ket{\widetilde{\Phi}}$ is entangled. The existence of such pairs are exactly the content of the concept of unconditional inseparability of superpositions, considered in the preceding section. It is always possible to find other states $\ket{\widetilde{\Psi}_2}$ and $\ket{\widetilde{\Psi}_3}$ such that the states $\ket{\widetilde{\Psi}_1}$, $\ket{\widetilde{\Psi}_2}$, $\ket{\widetilde{\Psi}_3}$, and $\ket{\widetilde{\Phi}}$ form a two-qubit orthonormal basis. See Refs.~\cite{Halder20, Halder21} for such bases. 

The states $\ket{\widetilde{\Psi}_1}$, $\ket{\widetilde{\Psi}_2}$, and $\ket{\widetilde{\Psi}_3}$ cannot be conclusively distinguished by LOCC. In particular, the state $\ket{\widetilde{\Psi}_1}$ cannot be conclusively identified by LOCC. This follows from the fact that if $\ket{\widetilde{\Psi}_1}$ can be conclusively identified by LOCC, then it is necessary and sufficient to find a product state which is nonorthogonal to $\ket{\widetilde{\Psi}_1}$ but orthogonal to $\ket{\widetilde{\Psi}_2}$ and $\ket{\widetilde{\Psi}_3}$ \cite{Chefles04, Bandyopadhyay09}. Obviously, this product state must belong to the subspace spanned by $\ket{\widetilde{\Psi}_1}$ and $\ket{\widetilde{\Phi}}$. But according to our assumption about the \(\{|\widetilde{\Psi}_1\rangle,|\widetilde{\Phi}\rangle\}\) pair, this subspace contains only one product state, viz. $\ket{\widetilde{\Phi}}$, because all others are  of the form, $a_1\ket{\widetilde{\Psi}_1}+a_2\ket{\widetilde{\Phi}}$, with \(a_1,a_2 \ne 0\), and  are entangled. And, $\ket{\widetilde{\Psi}_1}$ and $\ket{\widetilde{\Phi}}$ are orthogonal to each other. So, it is not possible to find a product state which is nonorthogonal to $\ket{\widetilde{\Psi}_1}$ but orthogonal to $\ket{\widetilde{\Psi}_2}$ and $\ket{\widetilde{\Psi}_3}$. It is important to mention here that any set of three orthogonal two-qubit maximally entangled states can always be conclusively distinguished by LOCC \cite{Halder21}. Therefore, the set consisting of the states \(\ket{\widetilde{\Psi}_1}\), \(\ket{\widetilde{\Psi}_2}\), and \(\ket{\widetilde{\Psi}_3}\) are ``more nonlocal'' than any set of three orthogonal two-qubit maximally entangled states, where the ``nonlocality'' is in the sense of conclusive local indistinguishability of a set of orthogonal shared quantum states. This also implies that the states \(\ket{\widetilde{\Psi}_1}\), \(\ket{\widetilde{\Psi}_2}\), and \(\ket{\widetilde{\Psi}_3}\) cannot be maximally entangled. This also follows from Ref.~\cite{Bravyi11}, because we have assumed that the state $\ket{\widetilde{\Phi}}$ is a product state and therefore, all of the states $\ket{\widetilde{\Psi}_1}$, $\ket{\widetilde{\Psi}_2}$, and $\ket{\widetilde{\Psi}_3}$ cannot be maximally entangled states.

It is clear that the average entanglement of the states of the set \(\{\ket{\widetilde{\Psi}_1}, \ket{\widetilde{\Psi}_2}, \ket{\widetilde{\Psi}_3}\}\) is lower than that of any set of maximally entangled two-qubit states. However, the states of the latter states are conclusively distinguishable by LOCC, while those of the former are not. We are therefore led to the phenomenon of {\it more nonlocality with less entanglement}~\cite{Horodecki03}. See Ref.~\cite{Halder21} in this regard. Locally indistinguishable mutually orthogonal ensembles of shared states can be used by a ``boss'' to send a secret to her ``sub-ordinates'' in such a way that they can recover the secret only if they cooperate. 

\subsection{Two-element ensembles and conclusive local indistinguishability}
Two orthogonal states of two qubits form the most elementary class of quantum ensembles. Such ensembles, if the constituent elements are pure, can never lead to local indistinguishability, perfect or conclusive~\cite{Walgate00}, leading to their uselessness from this perspective. To search for two-element ensembles of two qubits that are conclusively indistinguishable under LOCC, at least one of the two elements must be a mixed state. If we further require that the sum of the dimensions of the supports of the states within a two-element ensemble must not be equal to the total dimension of the Hilbert space, then in case of two qubits, the only option is to  consider ensembles consisting of a mixed state of rank-2 and a pure state. Consider now the pure state $\ket{\widetilde{\Psi}_1}$ and the mixed state  $\widetilde{\varrho}=p_1\ket{\widetilde{\Psi}_2}\bra{\widetilde{\Psi}_2}+p_2\ket{\widetilde{\Psi}_3}\bra{\widetilde{\Psi}_3}$, where $p_1, p_2>0$ and $p_1+p_2=1$. The states \(|\widetilde{\Psi}_1\rangle\), \(|\widetilde{\Psi}_2\rangle\), and \(|\widetilde{\Psi}_3\rangle\) are exactly as in the preceding subsection. As long as $\ket{\widetilde{\Psi}_1}$ cannot be conclusively identified by LOCC for the set, \(\{|\widetilde{\Psi}_1\rangle, |\widetilde{\Psi}_2\rangle, |\widetilde{\Psi}_3\rangle\}\), the two-element ensemble, \(\{|\widetilde{\Psi}_1\rangle\langle \widetilde{\Psi}_1|, \widetilde{\varrho}\}\), is also not conclusively distinguishable by LOCC. See Ref.~\cite{Halder20} in this regard. 

\section{Conclusion}\label{sec4}
Before we summarize our findings, we mention that the quantitative bounds on entanglement of superpositions that are already known in the literature might not be always useful to identify if the state after a superposition is separable or entangled. An example will be helpful to understand this. Let us consider an entangled state $\ket{\phi_1} = a\ket{00}+b\ket{11}+c\ket{22}$, $0<a,b,c<1$, $a^2+b^2+c^2=1$ and a product state $\ket{\phi_2} = \ket{01}$. These two states are orthogonal to each other. And, no nontrivial superposition of these two states can produce a product state, i.e., the state $\ket{\psi} = \alpha_1\ket{\phi_1}+\alpha_2\ket{\phi_2}$ is always entangled for $0<\alpha_1, \alpha_2<1$, $\alpha_1^2+\alpha_2^2=1$~\footnote{Although \(\alpha_1\) and \(\alpha_2\) are chosen here as real, picking complex values for them does not change the conclusion. This can be seen as follows. Clearly,  it is enough to have an extra phase attached to only one (any one) among \(\alpha_1\), \(\alpha_2\), and let us  choose it  to be \(\alpha_2\). Now this phase can be pulled into the definition of the \(|0\rangle\) of the first party. However, then an extra phase appears in the term \(\alpha_1 a |00\rangle\), which can then be absorbed in the \(|0\rangle\) of the second party. These re-definitions of the local bases are local unitaries, which does not affect the entanglement content of \(|\psi\rangle\).}. Notice that for this example, to detect entanglement, we need an inequality $E(\ket{\psi})>0$ and this must hold for all values of $a,b,c,\alpha_1,\alpha_2$ as defined above. It is actually difficult to find such a strict inequality. For example, one can try to use the lower bounds on the entanglement of superpositions, reported so far (such as Theorem 5 of \cite{Gour07}, Theorem 2 of \cite{Yu07}, Theorem 3 of \cite{Niset07}, or, Theorem 3 of \cite{Akhtarshenas11}). It is easy to check that using these bounds, it is not possible to prove that  entanglement of the state \(\ket{\psi}\) is  nonzero for all values of $0<a,b,c<1$, $a^2+b^2+c^2=1$ and $0<\alpha_1,\alpha_2<1$, $\alpha_1^2+\alpha_2^2=1$. But Theorem 1 of this paper can serve the purpose of detecting entanglement in $\ket{\psi}$. 

Usually, the quantitative bounds which are reported in the previous papers, contain a term related to the entanglement of the output state. Now, if we calculate the amount of entanglement of the output state, then obviously, we can tell whether the output state is separable or entangled. But in our case, it is not required to calculate the entanglement of the output state, e.g. for concluding if the output state is an entangled state, as for instance seen in Theorem 1 of the present paper. Of course, the quantitative results in the literature are very useful, and moreover, the qualitative results presented here are also not conclusive in all cases. But what we wish to argue is that both quantitative and qualitative analyses could be of value in unknotting the status of entanglement and separability in superpositions of quantum states. 

To conclude, we have explored separability and inseparability of superpositions of quantum states of shared systems. We have introduced the notions of  conditional and unconditional inseparability of superpositions. Specifically, we have shown that all nontrivial  superpositions of all pairs of an entangled state and a product state are entangled for bipartite systems, when the initial entangled state is of Schmidt rank three or higher. Obviously, the two-qubit system is left out in this result, and it is subsequently analyzed in detail, and we provided several specific cases there.  We then found that the considerations are useful in several quantum communication tasks. In particular, we identified a class of unextendible entangled bases, which we referred to as \(r\)-UEBs, and proved that they can contain no state which is conclusively locally identifiable. Moreover, the notion of unconditional inseparability of superpositions is found useful to exhibit the phenomenon of more nonlocality with less entanglement in a two-qubit system. Furthermore, we  have established a one-to-one correspondence between the phenomenon {\it more nonlocality with less entanglement} and a class of two-element ensembles which cannot be conclusively distinguished by local quantum  operations and classical communication. 

\section*{Acknowledgment}
We acknowledge support from the Department of Science and Technology, Government of India through the QuEST grant (Grant No.~DST/ICPS/QUST/Theme-3/2019/120).

\bibliography{ref}
\end{document}